# Pauli-limiting effects in the upper critical fields of a clean LiFeAs single crystal

Seunghyun Khim[1], Bumsung Lee[1], Jae Wook Kim[1], Eun Sang Choi[2], G. R. Stewart[3], and Kee Hoon Kim[1*]

[1]*CeNSCMR, Department of Physics and Astronomy, Seoul National University, Seoul 151-747, Republic of Korea*

[2]*National High Magnetic Field Laboratory, Florida State University, Tallahassee, Florida 32310, USA*

[3]*Department of Physics, University of Florida, Gainesville, Florida 32611-8440, USA*

## ABSTRACT

We have investigated the temperature-dependence of the upper critical field $H_{c2}(T)$ in a LiFeAs single crystal by direct measurements of resistivity under static magnetic fields up to 36 T. We find in the case of a magnetic field $H$ along the *ab*-plane that $H_{c2}^{ab}(0) = 30$ T is clearly lower than the orbital limiting field $H_{c2}^{orb,ab}(0) = 39.6$ T estimated by the $|dH_{c2}^{ab} / dT|_{Tc}$, suggesting the presence of both Pauli- and orbital-limiting effects in the pair breaking process. The best fit of $H_{c2}^{ab}(T)$ to the Werthamer-Helfand-Hohenberg formula results in the Maki parameter $\alpha = 0.9$ and negligible spin-orbit scattering constant ($\lambda_{so} = 0.0$). On the other hand, for $H$ along the *c*-axis, $H_{c2}^{c}(T)$ increases linearly down to our lowest temperature of 0.8 K, which can be explained by the multi-band effects. The anisotropy ratio $H_{c2}^{ab}(T) / H_{c2}^{c}(T)$ is 3 near $T_c$ and systematically decreases upon lowering temperature to become 1.3 at zero temperature. A comparative overview of the behavior of $H_{c2}^{ab}(T)$ in various Fe-based superconductors shows that, similar to LiFeAs, the calculated $H_{c2}^{orb,ab}(0)$ is generally much larger than the measured $H_{c2}^{ab}(0)$ and thus finite $\alpha$ values ranging from ~ 0.4 to 3 are necessary to describe the low temperature $H_{c2}^{ab}(T)$ behaviors. Moreover, LiFeAs is found to have the smallest $|dH_{c2}^{ab} / dT|_{Tc}$ values, indicating that LiFeAs is one of the cleanest Fe-based superconductors with a finite Maki parameter. We also discuss the implications of multi-band effects and spin-orbit scattering based on the finding that the estimated Pauli-limiting field is generally much larger than the BCS prediction in the Fe-based superconductors.

*corresponding author: khkim@phya.snu.ac.kr.

# I. INTRODUCTION

## A. Upper critical fields and pair-breaking mechanism

The discovery[1] of Fe-based superconductors has triggered enormous research efforts[2-3] in recent years, with the motivations, for example, to find a higher temperature superconductor and to understand the perhaps unconventional pairing mechanism. Measurement of the upper critical field $H_{c2}$ is an important part of this effort, since it can give clues (for a discussion, see ref. 3) to understanding various superconducting properties such as coherence lengths, coupling strength and pair-breaking mechanism. The anisotropy of $H_{c2}$, which is related to the dimensionality and the topology of the underlying electronic structure, also becomes important for superconducting wire applications as well as for understanding multi-band effects.

Generally, there exist two distinct ways to induce pair-breaking in type-II superconductors by an applied magnetic field, i.e., orbital and spin-paramagnetic effects. The former is related to an emergence of Abrikosov vortex lines and superconducting currents around vortex cores, which then reduce the condensation energy. The orbital limiting field refers to the critical field at which vortex cores begin to overlap and is given as $H_{c2}^{orb} = \Phi_0 / 2\pi\xi^2$ where $\xi$ is the coherence length and $\Phi_0 = 2.07 \times 10^{-15}$ T m$^2$ is the flux quantum. For one-band BCS superconductors, $H_{c2}^{orb}(0)$ is commonly derived from the slope of the determined $H$-$T$ phase boundary at $T_c$, which is given as $H_{c2}^{orb}(0) = -0.69|dH_{c2}/dT|_{T_c} T_c$ in the dirty limit and $-0.73|dH_{c2}/dT|_{T_c} T_c$ in the clean limit.[4]

The spin-paramagnetic pair-breaking effect comes from the Zeeman splitting of spin singlet Cooper pairs. The Pauli-limiting field $H_P$ is derived from the condition that the Zeeman energy in the normal state compensates the superconducting condensation energy under magnetic fields, i.e., $(1/2)\chi_N H_P^2 = (1/2)N(E_F)\Delta^2$, which yields $H_P(0) = g^{-1/2}\Delta / \mu_B$ (Chandrasekhar-Clogston limit).[5-6] Here, $\chi_N = g\mu_B^2 N(E_F)$ is the normal state spin susceptibility, $g$ is the Lande $g$-factor, $\mu_B$ is the Bohr magneton, $\Delta$ is the superconducting gap and $N(E_F)$ is the density of state at the Fermi level $E_F$. For a BCS superconductor where $2\Delta(0) = 3.52 k_B T_c$, $H_P(0)$ becomes $H_P^{BCS}(0) = 1.84 T_c$.

The actual $H_{c2}$ of real materials is generally influenced by the both orbital and spin-paramagnetic effects. The relative importance of the orbital and spin-paramagnetic effects can be described by the Maki parameter,[7]

$$\alpha = \sqrt{2}\,\frac{H_{c2}^{orb}(0)}{H_P(0)} \qquad (1)$$

Since $\alpha$ is known to be the order of $\Delta(0) / E_F$, $\alpha$ is usually $<< 1$. However, in materials with a heavy electron mass or multiple small Fermi pockets, $E_F$ can become quite small to result in $\alpha \geq 1$, yielding a possibility of realizing the Fulde-Ferrel-Larkin-Ovshnikov (FFLO) state, in which inhomogeneous superconducting phases with a spatially modulated order parameter and spin polarization is stabilized.[8-9]

In the Fe-based superconductors, in which five $d$-orbitals can contribute to the Fermi surfaces, multi-band effects should be also considered in the orbital-limiting mechanism. Based on the experimental studies of the well-known two band superconductor $MgB_2$, the multi-band effects are expected to result in either quite linear or even upward curvature in the $H_{c2}(T)$ curves near $T_c$.[10] Moreover, the anisotropy ratio $\gamma_H = H_{c2}^{ab}(T) / H_{c2}^{c}(T)$ shows strong temperature-dependence, in contrast to the temperature-independent behavior expected in a one-band superconductor.

B. Overview of $H_{c2}$ studies in Fe-based superconductors

A clear upward curvature was found in the $H_{c2}^{c}(T)$ of the '1111' system, $Re$FeAsO ($Re$ = rare earth) near $T_c$, supporting the presence of the multi-band effects.[11-14] For the '122' systems such as $A$Fe$_2$As$_2$ ($A$ = Ba, Sr, Ca, and Eu), several experimental studies also reported that $H_{c2}^{c}(T)$ exhibits quite a linear increase down to lowest temperatures and $\gamma_H$ is ~ 2 - 4 near $T_c$ and reduces toward 1 at low temperatures possibly due to the band warping effects.[15-18] On the other hand, experimental evidence for the spin-paramagnetic effects are also accumulating in the Fe-based superconductors. An oxygen deficient LaFeAsO showed a steep increase in $H_{c2}(T)$ near $T_c$, followed by a saturation behavior around $T_c / 2$, suggesting a strong spin-paramagnetic effect.[19-20] We also found clear evidence for a dominant Pauli-limited $H_{c2}$ behavior in a stoichiometric '11' system (Fe(Se,Te)); both $H_{c2}^{c}$ and $H_{c2}^{ab}$ show clear saturation at temperatures not far from $T_c$, resulting in $H_{c2}^{ab}(0) \approx H_{c2}^{c}(0) \approx 48$ T.[21-23] Moreover, we have recently suggested that the spin-paramagnetic pair-breaking effect should be also considered in the '122' system, based on the observations of robust pseudo-isotropic $H_{c2}(0)$ behaviors and the flattening in the $H_{c2}^{ab}(T)$ curve at low temperatures in a broad doping range and in various forms of the '122' materials.[18] Therefore, it is of prime importance at this stage to check whether the other Fe-based superconductors are also subject to such mixed pair-breaking processes, i.e. both multi-band-orbital- and Pauli-limiting effects.

## C. Characteristics and previous $H_{c2}$ studies of LiFeAs

LiFeAs, a representative compound in the '111' system, is regarded as a unique Fe-based superconductor because it shows $T_c \approx 18$ K without any nominal impurity or carrier doping.[24-27] By virtue of the minimal impurity or disorder effects, LiFeAs shows a very small residual resistivity $\rho_0$ = 4 - 20 μΩ cm and a large room temperature and residual resistivity ratio (RRR), $\rho(300\ K) / \rho_0 =$ ~ 20 - 65.[27-29] This is the second largest RRR value found in the Fe-based superconductors after KFe$_2$As$_2$ in which the RRR = ~ 100 - 1000 was observed.[30-32] The mean free path expected from the small residual resistivity is also significantly longer than the superconducting coherence length, supporting the idea that the system is a clean-limit superconductor (*vide infra*).

Partly due to the difficulty in preparing electrical contacts on a highly hygroscopic LiFeAs, the $H_{c2}$ studies based on the contactless methods were firstly reported; a torque magnetometer study[33] and a tunnel diode resonator (TDR) measurement.[34] The $H_{c2}^{ab}(0)$ ($H_{c2}^{c}(0)$) values are similar but detailed temperature-dependence shows a significant difference. We also note that the samples used for those previous works were based on different growth techniques, i.e., the Bridgman method for the TDR study and a self-flux method for the torque magnetometry. It is thus worthwhile to study $H_{c2}$ of the LiFeAs crystals from various sample growth techniques and extract their intrinsic, reproducible behavior to draw a common physical picture.

Herein, we study the temperature-dependent upper critical field $H_{c2}(T)$ of LiFeAs single crystals grown by a Sn-flux method as determined by direct dc resistivity measurements under static, high magnetic fields up to 36 T. $H_{c2}^{ab}(0)$ and $H_{c2}^{c}(0)$ were found to be 30 and 24 T, respectively. Based on a fit to the Werthamer-Helfand-Hohenberg (WHH) model, we identify the Pauli-limited $H_{c2}^{ab}(T)$ behavior with the Maki parameter $\alpha = 0.9$, while we need to apply the two-band model to explain its linearly increasing behavior for the $H_{c2}^{c}(T)$ data. Comparison of the $H_{c2}$ behavior of LiFeAs with other Fe-based superconductors implies that LiFeAs is a clean superconductor subject to both orbital and spin-paramagnetic pair-breaking effects.

## II. EXPERIMENTS

High quality single crystalline LiFeAs was grown by the Sn-flux method as reported earlier by our group.[27] Stoichiometric amounts of chemical elements with the Sn-flux was mixed with the ratio of [LiFeAs] : Sn = 1 : 10 in an alumina crucible and sealed in an evacuated quartz ampoule filled with a partial atmosphere of Ar gas. The ampoule was heated up to 850 °C and slowly cooled down to 500 °C. We used a centrifuge to separate the crystals from the molten flux. Shiny plate-like single crystals were obtained with a typical lateral area of 5 x 5 mm². Resistivity measurement was done with a conventional 4-probe method inside a ³He cryostat down to 0.8 K. We have confirmed through repeated growth efforts that the Sn-flux method provides very reproducible high quality single crystals, in which a bulk superconductivity is evidenced by a heat capacity jump and dc resistivity shows a sharp superconducting transition width $\Delta T_c$ = 1.1 K and a large RRR value (18 - 25) comparable to the samples growth by Bridgman or self-flux methods.[26,29] The successful growth of LiFeAs by the Sn-flux method should be therefore distinguished from the case of the Sn-flux grown BaFe$_2$As$_2$ where high quality specimens could not be well obtained.[35] Due to the hygroscopic nature of LiFeAs, the sample was covered with a Stycast™ epoxy after making electric contacts with a silver epoxy (Epotek™) to the single crystal's surface inside the glove box. Based on the fact that the superconducting transition width of 2.7 K is much larger than the as-grown sample, there may have been some degradation in the sample inside the Stycast™ and during the extended time of about 1 month before measurements at National High Magnetic Field Laboratory (NHMFL). However, the high field transport properties reported here are quite consistent with the as-grown sample in a low field region and thus appear to be intrinsic. Static magnetic field was applied up to 36 T along the *ab*-plane and *c*-axis directions by a resistive magnet at NHMFL in Tallahassee, USA.

## III. RESULTS

Fig. 1 shows the temperature-dependence of the in-plane resistivity. The resistivity monotonically decreased with decreasing temperature and showed no anomaly down to $T_c$ from room temperature (the inset). The superconducting transition temperature is estimated as $T_c^{50\%}$ = 17.4 K and transition width $\Delta T_c$ defined as $T_c^{90\%}$ - $T_c^{10\%}$ is 2.7 K, which is somewhat broader than $\Delta T_c$ = 1.1 K ($T_c^{50\%}$ = 17.4 K) in our reported, Sn-flux grown sample. As mentioned, a small degradation of the sample quality inside the Stycast™ epoxy is thought to cause this broader transition. Upon linearly extrapolating the $\rho$ data in 36 T (solid circles), we obtain residual resistivity $\rho_0$ = 29 μΩ cm while a fit to the normal state resistivity via $\rho(T) = \rho_0 + AT^2$ provides a bit higher $\rho_0$ = 32 μΩ cm; these results predict the RRR, as 20 and 18.4, respectively. These RRR values are consistent with that of a Sn-flux grown crystal reported previously (~ 24)[27] although it is somewhat smaller than those grown by a self-flux (~ 38)[29] or by a Bridgman technique (~ 45).[26]

To determine the temperature-dependence of $H_{c2}$, we measured isothermal resistivity vs. $H$ curves at selected temperatures from 0.8 to 16 K. Fig. 2 (a) and (b) show the results for the *ab*-plane ($H // ab$) and *c*-axis ($H // c$), respectively, in which the resistivity changes from zero to a finite value due to the suppression of superconductivity as $H$ increases through $H_{c2}$ at each temperature. The zero resistivity state, i.e., superconductivity is maintained up to higher fields for $H // ab$ than for $H // c$ while the transition width becomes broader for $H // ab$, presumably due to a weaker vortex pinning. Although only the up-sweeps are plotted in the Fig. 2, the field-up- and down-sweeps produced almost identical curves. Based on these data in Figs. 2(a) and (b), we determined $H_{c2}$ at each temperature for both directions with a criterion that 50 % of the normal state resistivity is realized at $H_{c2}$.

Fig. 3 shows the thus determined $H_{c2}(T)$ data for both $H // ab$ and $H // c$ directions (solid symbols) along with the low field $H_{c2}(T)$ data up to $H = 9$ T measured for a different piece of crystal from the same batch (open symbols).[27] The high field data follow well the curvature and the values of the low field ones, showing that the two samples used for the high and low field experiments produce consistent results each other. Upon decreasing temperature near $T_c$, $H_{c2}$ curves for both directions increase linearly with the slopes, $(dH_{c2} / dT)_{T_c}$ = -3.30 and -1.64 T / K for $H // ab$ and $H // c$, respectively. The orbital limiting field is predicted as $H_{c2}^{orb}(0) = 39.6$ T (19.7 T) for $H // ab$ ($H // c$) in the dirty limit and as 41.9 T (20.8 T) for $H // ab$ ($H // c$) in the clean limit. The corresponding Ginzburg-Landau coherence length in the dirty limit is then obtained as $\xi_{ab}$ = 40.9 Å and $\xi_c$ = 20.3 Å, and the coherence lengths are expected to be shorter in the clean limit, $\xi_{ab}$ = 39.8 Å and $\xi_c$ = 19.8 Å.

## IV. DISCUSSION

### A. Application of the WHH model

The temperature-dependence of $H_{c2}$ determined by the orbital and spin-paramagnetic effect in one band, dirty-limit superconductors is given by the WHH formula,

$$\ln\frac{1}{t} = \sum_{\nu=-\infty}^{\infty}\left(\frac{1}{|2\nu+1|} - \left[|2\nu+1| + \frac{\hbar}{t} + \frac{(\alpha\hbar/t)^2}{|2\nu+1| + (\hbar+\lambda_{so})/t}\right]^{-1}\right) \quad (2)$$

where $t = T/T_c$, $\hbar = (4/\pi^2)(H_{c2}(T)/|dH_{c2}/dT|_{Tc})$, $\alpha$ is the Maki parameter, and $\lambda_{so}$ is the spin-orbit scattering constant.[36] When $\lambda_{so} = 0$, $H_{c2}(0)$ obtained from the WHH formula satisfies the relation,

$$H_{c2}(0) = \frac{H_{c2}^{orb}(0)}{\sqrt{1+\alpha^2}} \quad (3)$$

which is originally derived by K. Maki.[7] In Fig. 3, we note that the experimental $H_{c2}$ curves for both $H$ directions significantly deviate from the predictions of the WHH model considering the orbital pair-breaking only, i.e., $\alpha = 0$ (dashed lines). For the $H // ab$ direction, the experimental $H_{c2}^{ab}$ curve exhibits a clear flattening at low temperatures compared to the expected $H_{c2}^{orb}(0)$ with $\alpha = 0$. Therefore, to describe the flattened $H_{c2}^{ab}(T)$ shape, we need to consider the Pauli-limiting effect as well by turning on a finite $\alpha$ and the best fit was obtained with $\alpha = 0.9$. It is noteworthy here that the spin-orbit scattering was not necessary to have the best fit ($\lambda_{so} = 0$). The obtained best fit parameters of $\alpha = 0.9$ and $\lambda_{so} = 0$ should be thus distinguished from other recently reported results, in which the application of the same WHH model produced a bit larger $\alpha = 1.74$ and 2.30 with relatively large spin-orbital scattering $\lambda_{so} = 0.3$ and 0.51, respectively.[33,37]

B. Multi-band effects and the $H_{c2}$ anisotropy

Fig. 3 shows that $H_{c2}^{c}$ increases all the way down to the lowest temperature 0.8 K so that the orbital WHH prediction, fitting fairly well near $T_c$, indeed underestimates the experimental $H_{c2}^{c}$ at low temperatures, i.e, $H_{c2}^{c}(0) > H_{c2}^{orb, c}(0)$. The quasi-linear increase in $H_{c2}^{c}(T)$ has been commonly observed in $MgB_2$ and other Fe-based superconductors ('122' and '1111')[11-18] and regarded as a hallmark of multi-band effects. The linearly increasing behavior has been successfully explained by an effective theoretical model considering only two main bands in the dirty limit for $MgB_2$ as well as '1111' system.[11,12,14] We could also show that our experimental $H_{c2}^{c}(T)$ curve can be successfully explained by the effective two-band model, clearly supporting that LiFeAs is a multi-band superconductor (not shown here). On the other hand, the fit based on the two-band model was not decisive in extracting the strength or the sign of intra- or inter-band coupling constants as similarly found in the '1111' system as well.

As the combined effects of flattened $H_{c2}^{ab}(T)$ and linearly increasing $H_{c2}^{c}(T)$, the $H_{c2}$ anisotropy $\gamma_H$ in LiFeAs exhibits strong temperature-dependence. As summarized in the inset of Fig. 3, $\gamma_H$ near $T_c$ is close to 3 but monotonically decreases to reach 1.3 as $T \rightarrow 0$ K. For $\gamma_H$ near $T_c$, the '122' system has resulted in values ranging from 1.5 to 4 and the '1111' system has shown larger $\gamma_H > 4$. Thus, in terms of the electronic structure related to the transport anisotropy, LiFeAs seems rather close to the '122' system.[11-18] On the other hand, the decrease in $\gamma_H(T)$ toward 1 at low temperatures has been similarly observed[3] in most of Fe-based superconductors to date. In some of the '122' materials, it was argued that the band warping effect can be an origin for the isotropic $\gamma_H(T)$ behavior.[15,17] However, the observation of the pseudo-isotropic $\gamma_H$ behavior in a wide class of Fe-based superconductors points to an alternative scenario that the Pauli-limiting effect can be the main origin.

Our results in this and previous sections show that $H_{c2}^{ab}(T)$ in LiFeAs can be explained by the one band WHH model considering both spin-paramagnetic and orbital pair breaking effects although the $H_{c2}^{c}(T)$ reflects clear multi-band effects. Based on the fact that application of $H // c$ rather than $H // ab$ is effective in forming closed orbits in cylindrical Fermi surfaces of LiFeAs,[38-39] we postulate that the multi-orbital effect becomes easily manifested in the $H_{c2}^{c}(T)$ curves while the multi-band model is not essential in explaining the shape of $H_{c2}^{ab}(T)$.

### C. Pauli-limiting effects in Fe-based superconductors

To understand better the interplay of orbital and Pauli-limiting effects in LiFeAs, we have tried to compare the experimental results in Fig. 3 with the published $H_{c2}(T)$ results in various Fe-based superconductors. Fig. 4 (a) compares the experimental $H_{c2}^{ab}(0)$ results with the predicted $H_{c2}^{\text{orb},ab}(0)$ by the WHH formula. Remarkably, we find that the actual $H_{c2}^{ab}(0)$ values (solid symbols) are significantly lower than $H_{c2}^{\text{orb},ab}(0)$ (open symbols) in most of the Fe-based superconductors. This suppression of the $H_{c2}^{ab}(T)$ curve at very low temperatures cannot be easily understood by the multi-band effect and directly supports that the Pauli-limiting plays an important role in determining the actual $H_{c2}^{ab}(0)$ in a broad class of Fe-based superconductors.

To describe this suppressed $H_{c2}^{ab}(0)$ behavior more quantitatively, we have herein resorted to the most well-established one band WHH formula. In this one band scheme, relative strength of the spin-paramagnetic effect over the orbital limiting effect can be simply understood by the magnitude of the Maki parameter $\alpha$. Therefore, we applied Eq. (3) to calculate $\alpha$ from the experimental $H_{c2}^{ab}(0)$ and $H_{c2}^{\text{orb},ab}(0)$ in various Fe-based superconductors. Note that the calculated $\alpha$ in this way does not necessarily include the effect of spin-orbit scattering. Fig. 4 (b) shows the calculated results; the '11' system shows the largest $\alpha \approx 3$, reflecting that spin-paramagnetic pair-breaking effect is dominant in determining $H_{c2}^{ab}(0)$ at low temperatures. On the other hand, the electron-doped '122' system shows relatively small $\alpha \approx 0.4$, indicating that the Paul-limiting becomes less important than the orbital-limiting effect. Compared with these two cases, $\alpha$ values of LiFeAs are scattered between 0.9 (present work) to 1.73, indicating that the Pauli-limiting effect is moderate. These $\alpha$ values are similar in magnitude to those of $LaO_{0.9}F_{0.1}FeAs_{1-\delta}$, holed-doped $Ba_{0.6}K_{0.4}Fe_2As_2$ and $KFe_2As_2$.[17,19,31] Because $\alpha$ values in LiFeAs are close to or even exceed unity, the '111' system, similar to the '11' system, might be another candidate to expect the FFLO ground state in Fe-based superconductors.

D. Cleanness of LiFeAs and initial slopes of $H_{c2}$ in Fe-based superconductors

To check the cleanness of the present LiFeAs, we compare the calculated coherent lengths in Section III with the mean free path in the normal state. Based on our *ab*-plane residual resistivity $\rho_0 = 29$ μΩ cm and the reported Hall coefficient $R_H = -2.7 \times 10^{-10}$ m$^3$ / C at 20 K by O. Heyer et al.,[29] the *ab*-plane mean free path is estimated as $l_{ab} = \hbar(3\pi^2)^{1/3} / e^2\rho_0 n^{2/3} = 52.5$ Å. This $l_{ab}$ is slightly larger than $\xi_{ab}$, indicating that our LiFeAs crystal is certainly not in the "dirty limit" ($l_{ab} \ll \xi_{ab}$) but closer to the clean limit ($l_{ab} \gg \xi_{ab}$). Yet another experimental parameter reflecting the carrier scattering is the initial slope of $H_{c2}$ near $T_c$. While the WHH model is based on a dirty-limit approximation, it relates $|dH_{c2} / dT|_{Tc}$ to the normal state resistivity $\rho_N$ and the density of states at the Fermi level $N(E_F)$ as $(4eck_B / \pi)N(E_F)\rho_N \sim (v_F l)^{-1}$, where $e$ is the charge of the electron, $c$ is the velocity of light, $k_B$ is the Boltzmann constant, $v_F$ is the Fermi velocity and $l$ is the mean free path. Thus, large $v_F l$ would result in smaller $|dH_{c2} / dT|_{Tc}$, presumably reflecting cleanness of a material through $|dH_{c2} / dT|_{Tc}$.

Fig. 5 summarizes $|dH_{c2} / dT|_{Tc}$ in various Fe-based superconductors including the present work on LiFeAs. We find that the '11' (Fe(Se,Te)) system particularly exhibits the largest $|dH_{c2}^{ab} / dT|_{Tc} \geq 9$ T / K and thus a large $H_{c2}^{orb, ab}(0)$ among Fe-based superconductors. Considering the realization of the large Maki parameter of $\alpha \approx 3$ in the '11' materials, it is most likely that the '11' system corresponds to a dominantly Paul-limited superconductor in the dirty limit. Previous observations of large spin-orbit scattering constant, small density of states, and large normal state resistivity all seem consistent with this postulate.[22] On the contrary, $|dH_{c2}^{ab} / dT|_{Tc}$ in the present LiFeAs crystal shows about 3 T / K, which is the smallest among Fe-based superconductors and comparable to that of KFe$_2$As$_2$ (3.8 T / K). It is also noted that the $|dH_{c2}^{ab} / dT|_{Tc}$ values in the published LiFeAs crystals are mostly located below 5 T / K, forming a group of samples with relatively small RRR values. Moreover, it is now well-known that KFe$_2$As$_2$ has the lowest normal state resistivity and the highest RRR values (> 1000) while the RRR of LiFeAs corresponds to the second largest. All these observations clearly support that LiFeAs is a clean superconductor being subject to both multi-band orbital- and Pauli-limiting effects.

It is interesting to find that the obtained Maki parameter $\alpha$ (0.9) is close to 1, which is a necessary condition for stabilizing the FFLO ground state in a clean limit superconductor. In a recent $H_{c2}(T)$ study for LiFeAs grown by the Bridgman technique, a clean limit theory was indeed applied to claim a possible realization of the FFLO state.[34] Our present findings in the $H_{c2}(T)$ behavior also supports that LiFeAs grown by the Sn-flux may be close to such an instability. Upon further tuning of the sample quality, it might be worthwhile to check the possible realization of the FFLO ground state by other experimental tools e.g., heat capacity and neutron scattering.

### E. Enhancement of the Pauli-limiting field

To estimate the actual Pauli-limiting field $H_P(0)$ in various Fe-based superconductors, we rewrite the $H_P(0)$ as

$$H_P(0) = \frac{\sqrt{2} H_{c2}^{orb}(0)}{\sqrt{(H_{c2}^{orb}(0)/H_{c2}(0))^2 - 1}} \qquad (4)$$

, which can be derived from Eqs. (1) and (3). Fig. 6 summarizes the calculated $H_P(0)$ in various Fe-based superconductors based on Eq. (4) and the experimental $H_{c2}^{ab}(0)$ data. We note that the $H_P(0)$ values thus obtained are overall 2 - 5 times larger than the $H_P^{BCS}(0) = 1.84 T_c$ in most of the Fe-based superconductors. In more detail, the $H_P(0)$ values of the LiFeAs system are found to be enhanced by 1.3 - 2 times of $H_P^{BCS}(0)$. For the '11' system, the $H_P(0)$ values are enhanced more to have ~ 2.6 times of $H_P^{BCS}(0)$. While the electron-doped (particularly Co-doped) '122' system shows a most enhanced $H_P(0)$, which corresponds to ~ 5 times, the hole doped $Ba_{0.6}K_{0.4}Fe_2As_2$ and $KFe_2As_2$ show 1.5 - 2 times of $H_P^{BCS}(0)$. Therefore, the degree of the enhancement seems to be the characteristics of each Fe-based superconductor system.

The expression of $H_P(0)$ in the Chandrasekhar-Clogston limit, i.e. $H_P(0) = g^{-1/2} \Delta / \mu_B$ shows that $H_P(0)$ can be enhanced by either strong coupling effect (i.e., $\Delta$ increases due to strong electron-boson coupling or strong correlation effect) or significant spin-orbit scattering (i.e., $\lambda_{so} \neq 0.0$, resulting in $g <$ 2).[40-41] Therefore, to properly understand the physical origin of the enhanced $H_P(0)$ over $H_P^{BCS}(0)$, it seems of importance to have reliable information on $\Delta$ and $\lambda_{so}$. Herein, we discuss the possible origin for the enhancement of $H_P(0)$ in each system based on the available $\Delta$ and $\lambda_{so}$ data.

It is worthwhile to reemphasize that LiFeAs is quite clean as compared with the other '11' or '122' systems as discussed in Section D. It is thus expected that the spin-orbit scattering is negligible in LiFeAs. Our fit results implying $\lambda_{so} = 0$ in Fig. 3 is also consistent with this reasoning. The strong coupling effect is then a most natural mechanism to explain the enhanced $H_P(0)$ in LiFeAs. Various investigations such as angle resolved photoemission spectroscopy (ARPES),[38-39] nuclear magnetic resonance (NMR),[42] lower critical field ($H_{c1}$),[43-44] and penetration depth measurements[28,45] have indeed shown that LiFeAs does not have a single gap but mainly two superconducting gaps, i.e., the small ($\Delta_s$) and the large gaps ($\Delta_L$). It is found from the literature[2-3] that $2\Delta_s$ is about 1.0 - 2.6 $k_BT_c$ and $2\Delta_L$ falls within the range of 3.6 - 6.0 $k_BT_c$. Thus, $2\Delta_L$ exceeds the BCS prediction of $2\Delta_{BCS} = 3.52$ $k_BT_c$ by a factor of 1 - 1.7. Therefore, the enhancement of the large gap can roughly explain its $H_P(0)$ enhancement factor (1.3 - 2) observed in LiFeAs (Fig. 6).

Our observation in LiFeAs further suggests that the other Fe-based superconductors may also show similar correlation in the enhancement factors of the $H_P(0)$ and $2\Delta_L$ over their BCS predictions. To check this, we have compiled available superconducting gap data in various Fe-based superconductors to extract their large gap value $\Delta_L$ among the observed multi-gap values and compare $\Delta_L / \Delta_{BCS}$ with $H_P(0) / H_P^{BCS}(0)$ in Fig. 7. In this effort, we did not include the experimental data in which only a single superconducting gap has been observed because it is uncertain whether the observed single gap can correspond to either the small or the large gap. In other words, we strictly focus on the cases where multiple gaps are confirmed in these supposedly multi-band Fe based superconductors and thereby mitigate the errors in the $\Delta_L$ estimation, which may come from the sample characteristics or the limits in the measurement techniques. Based on this criterion to plot Fig. 7, we discuss below the correlation between those two quantities and its implication in each Fe-based superconductor.

In the '11' system, the multi-gap feature was recently found to form $2\Delta_L / k_B T_c \approx 4 - 8.3$ and $2\Delta_S / k_B T_c \approx 0.8 - 4$ in optics,[46] $\mu$SR,[47] STS,[48] and specific heat studies.[49] It is thus important to note that the $2\Delta_L$ enhancement factor, corresponding to ~ 1 - 2.4 times of $2\Delta_{BCS}$, is generally larger than that of LiFeAs, indicating that the strong-coupling effect in the $2\Delta_L$ is more pronounced than the '111' case. This observation is consistent with the result in Fig. 7 that the enhancement factor of $H_P(0)$ in the '11' system is ~ 2.6 times of $H_P^{BCS}(0)$, which is clearly larger than that of LiFeAs (1.3 - 2). This finding thus supports again that the enhanced Pauli-limiting field in the '11' system can be linked to the enhancement of the large superconducting gap.[50] On the other hand, it should be noted that the '11' system apparently appears dirtier than the other Fe-based superconductors. For example, those observations of large normal state resistivity, small RRR values, and large $|dH_{c2}^{ab} / dT|_{Tc}$ constitute evidence for the dirtiness in the '11' system as compared with '111'. In this dirty limit, the spin-orbit scattering could not be overlooked and might also play a role in a pair breaking process in the '11' system. To be consistent with this scenario, in our previous work based on the WHH fitting, the finite $\lambda_{so}$ was essential to describe the curvature of $H_{c2}^{ab}(T)$ as well as $H_{c2}^{c}(0)$.[22-23] Therefore, in contrast to LiFeAs, we might need to include the spin-orbit scattering as an additional factor to enhance the $H_P(0)$ in this '11' system.

The correlation in the enhancement factors of the $H_P(0)$ and $2\Delta_L$ seems to exist in both '1111' and fully doped '122' systems as well. In the La(O,F)FeAs, the Andreev reflection study revealed two superconducting gaps as $2\Delta_L / k_B T_c \approx 6.4$ and $2\Delta_S / k_B T_c \approx 2.3$.[51] The ratio $\Delta_L / \Delta_{BCS} = 1.8$ are rather close to the gap enhancement factor $H_P(0) / H_P^{BCS}(0) = 2.2$ Moreover, in the fully hole-doped $KFe_2As_2$ system, an NMR measurement predicted $2\Delta_L / k_B T_c \approx 4.8$ and $2\Delta_S / k_B T_c \approx 0.60$.[52] Because the $\Delta_L$ is enhanced by 1.4 times of $\Delta_{BCS}$, the factor of $H_P(0) / H_P^{BCS}(0) = 1.5$ is quite consistent with the gap enhancement.

In both partially hole-doped $(Ba,K)Fe_2As_2$ and electron-doped '122' systems, the multi-gaps were observed and their magnitudes that can be roughly divided by two groups were indeed similar; the $2\Delta_L / k_B T_c \approx 7 - 10$ and $2\Delta_S / k_B T_c \approx 1.7 – 4.5$ were observed in ARPES,[53-55] Andreev reflection,[56] and STS measurements.[57] For $(Ba,K)Fe_2As_2$, the factor of $H_P(0)$ enhancement is 1.4 - 2 and this $H_P(0)$ value seems roughly consistent with the gap enhancement factor, 2 - 2.8. However in the electron-doped system, the $H_P(0)$ is enhanced by ~ 5 times of $H_P^{BCS}(0)$, which is clearly much bigger than the gap enhancement factor. We find in Fig. 5 that the electron-doped system has particularly small $\alpha$ =~ 0.4, showing that the effect of the spin-paramagnetic pair-breaking is rather small. In other words, the orbital pair breaking effect should be more important in the electron system than the hole-doped one. This observation implies that the correlation between the enhancement factors of $H_P(0)$ and $\Delta_L$ can becomes apparent only when the effect of the spin-paramagnetic pair-breaking becomes rather large. This is understandable because the proportionality relation between the Pauli-limiting field and the superconducting gap was indeed extracted by assuming only the Pauli-limiting effect without the orbital limiting effect. It is further worthwhile to mention that in the weak Pauli-limiting regime where the large value of $H_P(0) > 170$ T is predicted as in the electron-doped system, the actual $H_P(0)$ value would be very sensitive to a small variation of the Maki parameter. It is also suggested that the origin of the relatively weak Pauli-limiting effect in the electron-doped '122' system should be further understood based on e.g., its electronic structure or orbital character in the Fermi surface.

## IV. CONCLUSIONS

In summary, we have investigated the temperature-dependence of the upper critical fields in a clean LiFeAs single crystal under static magnetic fields up to 36 T. $H_{c2}^{ab}(T)$ show clear evidence of the presence of the Pauli-limiting effect at low temperatures. Applying the Werthamer-Helfand-Hohenberg model, we could extract a relatively large Maki parameter $\alpha = 0.9$ without spin-orbit scattering, which is close to the borderline to form the Fulde-Ferrel-Larkin-Ovshnikov ground state. Upon comparing $H_{c2}^{ab}(0)$ and $|dH_{c2}^{ab} / dT|_{T_c}$ in the literature, we conclude that LiFeAs is one of cleanest Fe-based superconductors being subject to the spin-paramagnetic pair-breaking along $H // ab$

1717

direction. We also showed that the estimated Pauli-limiting field is generally larger than the weakly coupled BCS prediction, and discussed strong coupling effects and spin-orbit scattering as its main origins.

## ACKNOWLEDGEMENT

This work was supported by the National Creative Research Initiatives (2010-0018300), National Research Laboratory program (M10600000238) by NRF, and the Fundamental R&D Program for Core Technology of Materials by MOKE, Korean Government. Experiments were performed at the National High Magnetic Field Laboratory, which is supported by the U.S. DOE, the National Science Foundation, and the State of Florida. The work in University of Florida is supported by the US Department of Energy, contract no. DE-FG02-86ER45268.

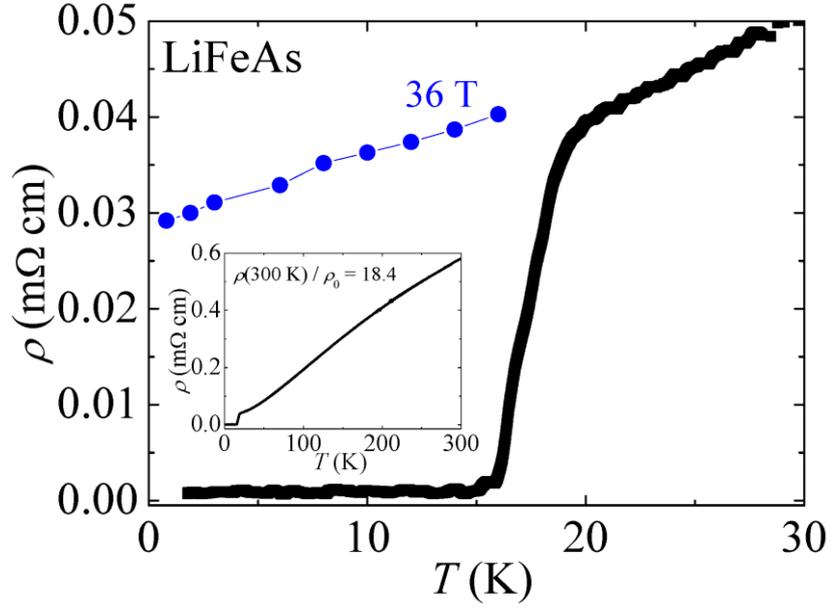

FIG. 1. (color online) Temperature dependence of in-plane resistivity of a LiFeAs single crystal grown by Sn-flux. The transition temperature is estimated as $T_c^{90\%}$ = 18.8 K, $T_c^{50\%}$ = 17.4 K and $T_c^{10\%}$ = 16.1 K, resulting $\Delta T_c = T_c^{90\%} - T_c^{10\%}$ = 2.7 K. The solid circles refer to the resistivity values at $H$ = 36 T, indicating that zero temperature resistivity $\rho_0$ is close to 29 $\mu\Omega$ cm as extrapolated by a linear fit. The inset shows the resistivity curve up to room temperature, giving a RRR value of 18.4.

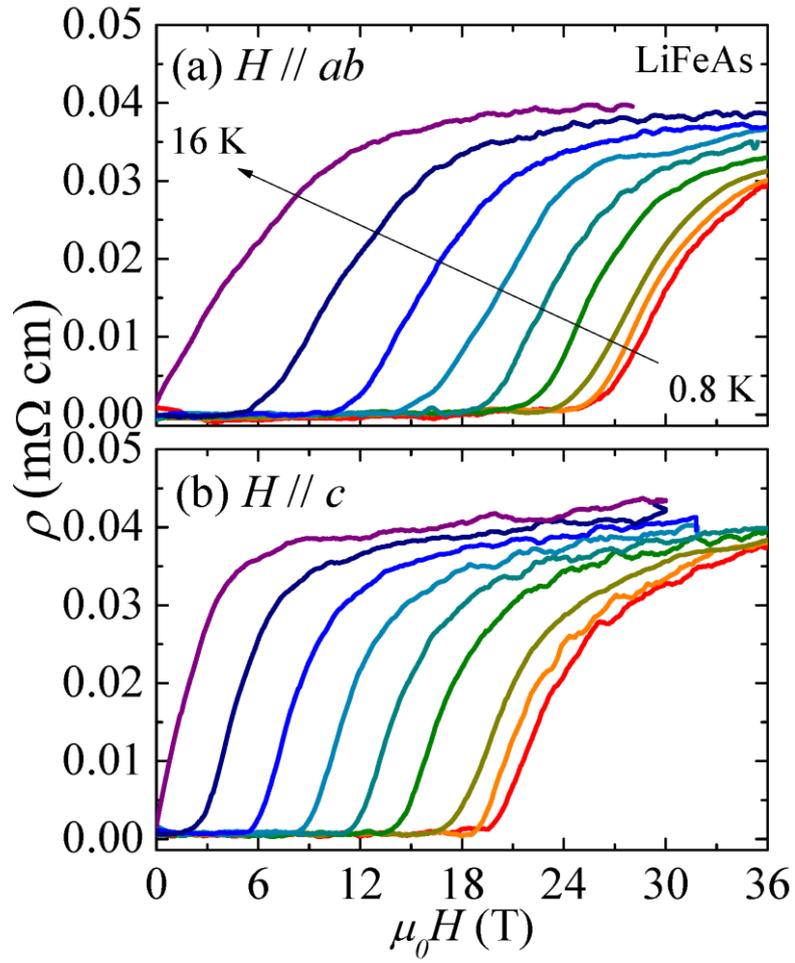

FIG. 2. (color online) Magnetic-field-dependence of resistivity of a LiFeAs single crystal at fixed temperatures; 0.8, 1.9, 3, 6, 8, 10, 12, 14 and 16 K. External magnetic field was applied along (a) *ab* – plane (*H // ab*) and (b) *c* - axis (*H // c*).

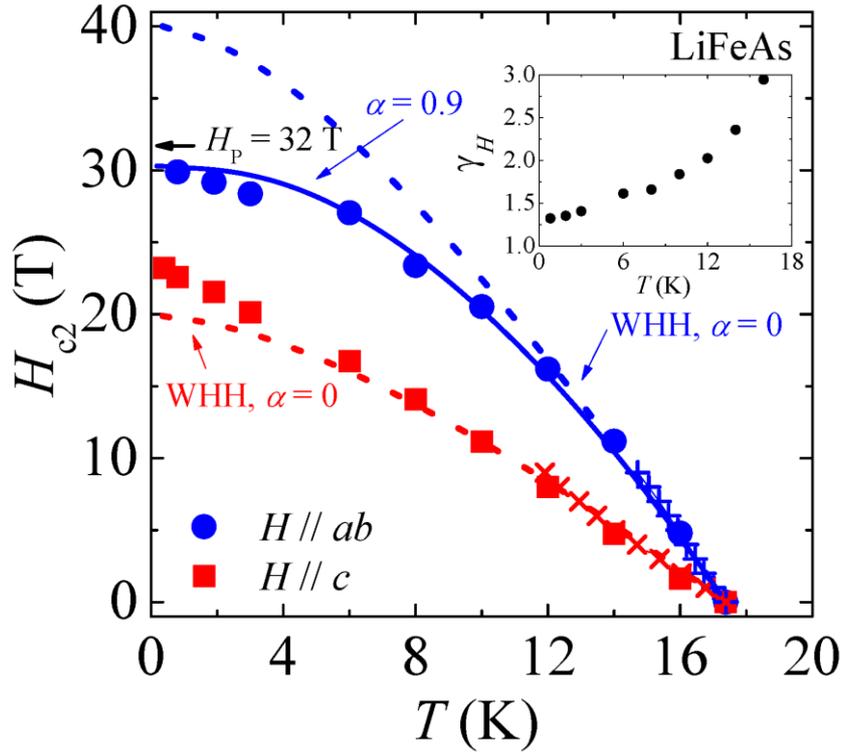

FIG 3. (color online) $H_{c2}(T)$ of LiFeAs for $H // ab$ and $H // c$ determined in this work (closed symbols). The open symbols represent $H_{c2}$ data from Ref. 27, which report low field experimental results for a different piece of single crystal from the same growth batch. The dotted lines are the WHH predictions with only an orbital pair-breaking effect included (i.e. the Maki parameter $\alpha = 0$) while the solid lines show the WHH fit with the Pauli-limiting effect considered (i.e., $\alpha = 0.9$). The inset shows the temperature-dependence of the anisotropy $\gamma_H = H_{c2}^{ab}(T) / H_{c2}^{c}(T)$.

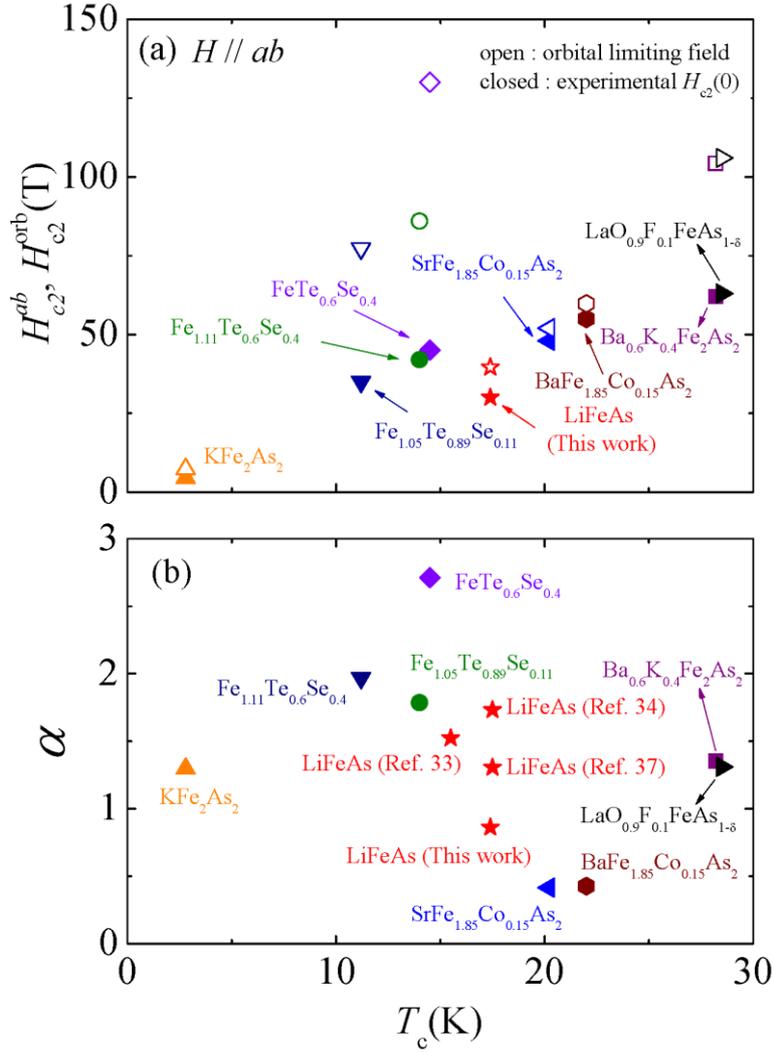

FIG. 4. (color online) (a) Summary of the measured $H_{c2}(0)$ (closed symbols) and $H_{c2}^{\text{orb}}(0)$ (open symbols) along $H \parallel ab$ and (b) the corresponding Maki parameters $\alpha = [(H_{c2}^{\text{orb},ab}(0) / H_{c2}(0))^2 - 1]^{0.5}$ in various Fe-based superconductors; $LaO_{0.9}F_{0.1}FeAs_{1-\delta}$,[19] $KFe_2As_2$,[31] $Fe_{1.11}(Te_{0.6}Se_{0.4})$,[21] $Fe(Te_{0.6},Se_{0.4})$,[22] $Fe_{1.05}(Te_{0.89}Se_{0.11})$,[23] $BaFe_{1.85}Co_{0.15}As_2$,[16] $(Ba_{0.6}K_{0.4})Fe_2As_2$,[7] $SrFe_{1.85},Co_{0.15}As_2$[8] and LiFeAs ; N. Kurita et al.,[33] K. Cho et al.,[34] J. L. Zhang et al.,[37] and this work.

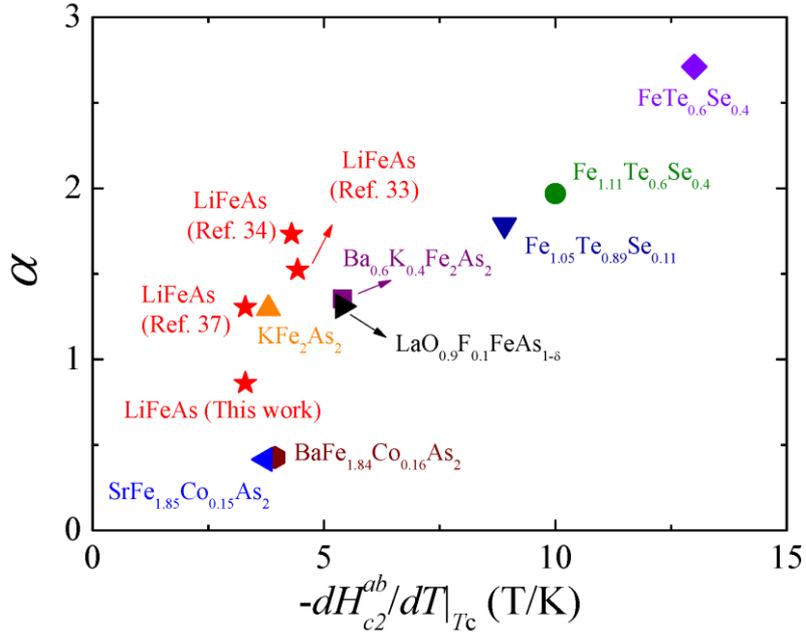

FIG. 5. (color online) The extracted Maki parameters with the initial slope of $H_{c2}$, $|dH_{c2}^{ab}/dT|_{T_c}$ for various Fe-based superconductors; LaO$_{0.9}$F$_{0.1}$FeAs$_{1-\delta}$,[19] KFe$_2$As$_2$,[31] Fe$_{1.11}$(Te$_{0.6}$Se$_{0.4}$),[21] Fe(Te$_{0.6}$,Se$_{0.4}$),[22] Fe$_{1.05}$(Te$_{0.89}$Se$_{0.11}$),[23] BaFe$_{1.85}$Co$_{0.15}$As$_2$,[16] (Ba$_{0.6}$K$_{0.4}$)Fe$_2$As$_2$,[7] SrFe$_{1.85}$,Co$_{0.15}$As$_2$,[8] and LiFeAs : N. Kurita *et al*.,[33] K. Cho *et al*.,[34] J. L. Zhang *et al*.,[37] and work done in this paper.

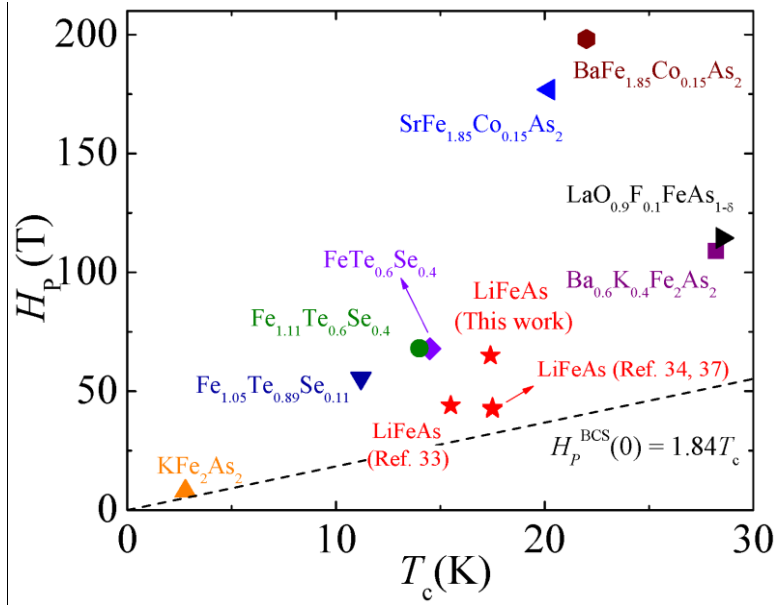

FIG. 6. (color online) The Pauli-limiting field $H_P(0)$ with $T_c$ for various Fe-based superconductors; $LaO_{0.9}F_{0.1}FeAs_{1-\delta}$,[19] $KFe_2As_2$,[31] $Fe_{1.11}(Te_{0.6}Se_{0.4})$,[21] $Fe(Te_{0.6},Se_{0.4})$,[22] $Fe_{1.05}(Te_{0.89}Se_{0.11})$,[23] $BaFe_{1.85}Co_{0.15}As_2$,[16] $(Ba_{0.6}K_{0.4})Fe_2As_2$,[17] $SrFe_{1.85},Co_{0.15}As_2$[8] and LiFeAs : N. Kurita *et al.*,[33] K. Cho *et al.*,[34] J. L. Zhang *et al.*,[37] and work done in this paper. [Note that the data from Ref. 34 and Ref. 37 is almost overlapped.] The dotted line shows the Pauli-limiting field for a weakly-coupled BCS superconductor.

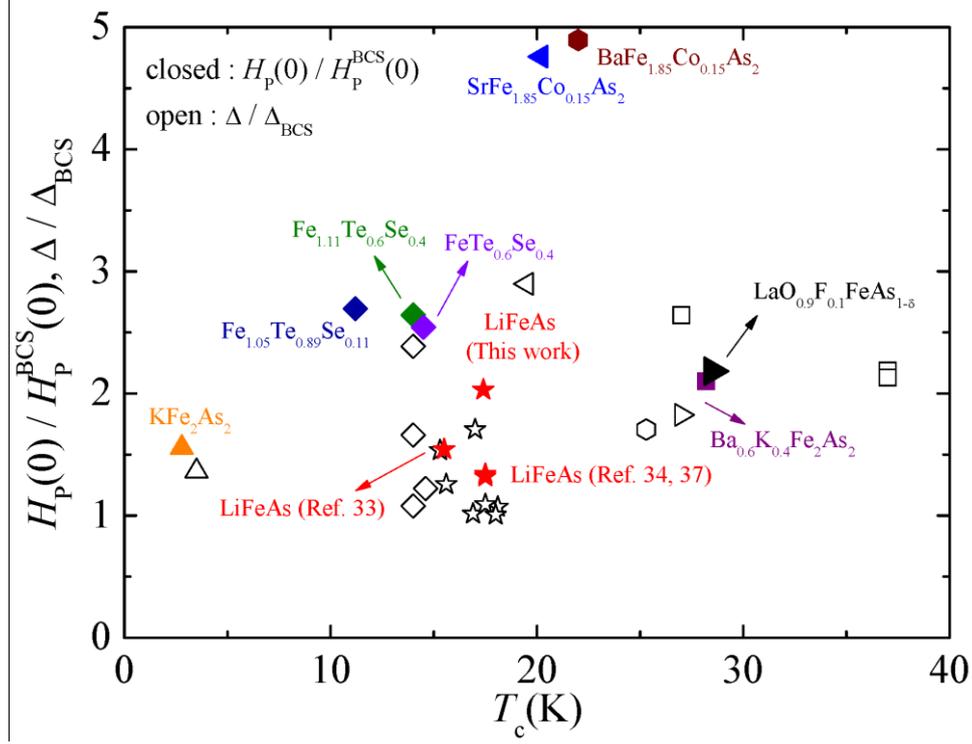

FIG. 7. (color online) The $H_P(0) / H^{BCS}_P(0)$ (closed symbols) with $T_c$ for various Fe-based superconductors including $KFe_2As_2$,[31] $Fe_{1.11}(Te_{0.6}Se_{0.4})$,[21] $Fe(Te_{0.6},Se_{0.4})$,[22] $Fe_{1.05}(Te_{0.89}Se_{0.11})$,[23] $BaFe_{1.85}Co_{0.15}As_2$,[16] $(Ba_{0.6}K_{0.4})Fe_2As_2$,[7] $SrFe_{1.85},Co_{0.15}As_2$[18] and LiFeAs. For the $H_P(0) / H^{BCS}_P(0)$ of LiFeAs, N. Kurita et al.,[33] K. Cho et al.,[34] J. L. Zhang et al.,[37] and work done in this paper were summarized. [Note that the data from Ref. 34 and Ref. 37 is almost overlapped.] Moreover, $\Delta / \Delta_{BCS}$ (open symbols) values were extracted from La(O,F)FeAs (Ref. 51), $KFe_2As_2$ (Ref. 52), Fe(Se,Te) (Refs., 48, 47, 49, and 46), LiFeAs (Refs. 38, 39, 28, 43, 45, 44, and 42), $Sr(Fe,Co)_2As_2$ (Ref. 57), $Ba(Fe,Co)_2As_2$ (Ref. 55) and $(Ba,K)Fe_2As_2$ (Refs. 54, 53, and 56). Note that the order of the multiple references in each compound is indeed proportional to the magnitude of the $\Delta / \Delta_{BCS}$.